\documentclass[11pt,a4paper]{article} 
\pdfoutput=1 

\usepackage{jcappub} 
\usepackage{lineno,hyperref}

\author[a]{C.~Deaconu}
\author[b]{L.~Batten}
\author[c]{P.~Allison}
\author[c]{O.~Banerjee}
\author[c]{J.~J.~Beatty}
\author[d]{K.~Belov}
\author[e,f]{D.~Z.~Besson}
\author[g]{W.~R.~Binns}
\author[g]{V.~Bugaev}
\author[h]{P.~Cao}
\author[i]{C.~H.~Chen}
\author[i]{P.~Chen}
\author[i]{Y.~Chen}
\author[h]{J.~M.~Clem}
\author[c]{A.~Connolly}
\author[j]{L.~Cremonesi}
\author[c]{B.~Dailey}
\author[g]{P.~F.~Dowkontt}
\author[k]{B.~D.~Fox}
\author[c]{J.~W.~H.~Gordon}
\author[k]{P.~W.~Gorham}
\author[l]{C.~Hast}
\author[k]{B.~Hill}
\author[i]{S.~Y.~Hsu}
\author[i]{J.~J.~Huang}
\author[a,c]{K.~Hughes}
\author[c]{R.~Hupe}
\author[g]{M.~H.~Israel}
\author[d]{K.~M.~Liewer}
\author[i]{T.~C.~Liu}
\author[a]{A.~B.~Ludwig}
\author[k]{L.~Macchiarulo}
\author[k]{S.~Matsuno}
\author[c]{K.~McBride}
\author[k]{C.~Miki}
\author[h]{K.~Mulrey}
\author[i]{J.~Nam}
\author[d]{C.~Naudet}
\author[b]{R.~J.~Nichol}
\author[e,f]{A.~Novikov}
\author[a]{E.~Oberla}
\author[c,e]{S.~Prohira}
\author[k]{R.~Prechelt}
\author[g]{B.~F.~Rauch}
\author[i]{J.~Ripa}
\author[k,m]{J.~M.~Roberts}
\author[d]{A.~Romero-Wolf}
\author[k]{B.~Rotter}
\author[k]{J.~W.~Russell}
\author[n]{D.~Saltzberg}
\author[h]{D.~Seckel}
\author[k]{H.~Schoorlemmer}
\author[i]{J.~Shiao}
\author[c]{S.~Stafford}
\author[e]{J.~Stockham}
\author[e]{M.~Stockham}
\author[n]{B.~Strutt}
\author[c]{M.~S.~Sutherland}
\author[k]{G.~S.~Varner}
\author[a]{A.~G.~Vieregg}
\author[n]{N.~Wang}
\author[i]{S.~H.~Wang}
\author[o]{S.~A.~Wissel}

\affiliation[a]{Dept. of Physics, Enrico Fermi Inst., Kavli Inst. for Cosmological Physics, Univ. of Chicago, Chicago, IL 60637.}
\affiliation[b]{Dept. of Physics and Astronomy, University College London, London, United Kingdom.}
\affiliation[c]{Dept. of Physics, Center for Cosmology and AstroParticle Physics, Ohio State Univ., Columbus, OH 43210.}
\affiliation[d]{Jet Propulsion Laboratory, California Institute for Technology,  Pasadena, CA 91109.}
\affiliation[e]{Dept. of Physics and Astronomy, Univ. of Kansas, Lawrence, KS 66045.}
\affiliation[f]{Moscow Engineering Physics Institute, Moscow, Russia.}
\affiliation[g]{Dept. of Physics, McDonnell Center for the Space Sciences, Washington Univ. in St. Louis, MO 63130.}
\affiliation[h]{Dept. of Physics, Univ. of Delaware, Newark, DE 19716.}
\affiliation[i]{Dept. of Physics, Grad. Inst. of Astrophys., Leung Center for Cosmology and Particle Astrophysics, National Taiwan University, Taipei, Taiwan.}
\affiliation[j]{School of Physics and Astronomy, Queen Mary University of London, London, United Kingdom.}
\affiliation[k]{Dept. of Physics and Astronomy, Univ. of Hawaii, Manoa, HI 96822.}
\affiliation[l]{SLAC National Accelerator Laboratory, Menlo Park, CA, 94025.}
\affiliation[m]{Center for Astrophysics and Space Sciences, Univ. of California, San Diego, La Jolla, CA 92093.}
\affiliation[n]{Dept. of Physics and Astronomy, Univ. of California, Los Angeles, Los Angeles, CA 90095.}
\affiliation[o]{Physics Dept., Pennsylvania State Univ., State College, PA 16802.}

\collaboration{ANITA Collaboration}

 \emailAdd{cozzyd@kicp.uchicago.edu} 

\arxivnumber{2010.02869}

\newcommand{\txs}{TXS 0506+056}
\newcommand{\icemc}{\texttt{icemc}}

\title{A search for ultrahigh-energy neutrinos associated with astrophysical\\ sources using the third flight of ANITA}

\abstract
  {
The ANtarctic Impulsive Transient Antenna (ANITA) long-duration balloon
  experiment is sensitive to interactions of ultrahigh-energy ($E>10^{18}$ eV)
  neutrinos in the Antarctic ice sheet. The third flight of ANITA, lasting 22
  days,  began in December 2014. We develop a methodology to search for
  energetic neutrinos spatially and temporally coincident with potential source
  classes in ANITA data. This methodology is applied to several source classes:
  the potential IceCube-identified neutrino sources \txs and NGC 1068,  flaring high-energy blazars reported by the Fermi
  All-Sky Variability Analysis, gamma-ray bursts, and supernovae. Among 
  searches within the five source classes, one candidate was identified as associated
  with SN 2015D, although not at a statistically significant level. We proceed to place upper limits on the source classes. We further
  comment on potential application of this methodology to more
  sensitive future instruments.
}

\begin{document}
\maketitle

\section{Introduction} 

The Antarctic Impulsive Transient Antenna (ANITA)
experiment~\citep{anitaInstrument} deploys a balloon-borne radio interferometer
to search for the impulsive Askaryan radio emission~\citep{askaryan,slac_ice}
expected to be produced by the interactions of ultrahigh-energy (UHE) neutrinos
($E>10^{18}$ eV) interacting in polar ice.  ANITA has previously reported
constraints on diffuse UHE neutrinos~\citep{anita1,anita2,anita3,anita4} as well
as neutrinos in time-coincidence with gamma-ray bursts (GRBs)~\citep{anita2_grb}.
No candidate events have been observed above background expectations so far in the Askaryan channel,
but ANITA sets the most stringent limits on diffuse UHE neutrino flux 
above $10^{19.5}$ eV.

\emph{Cosmogenic} UHE neutrinos are expected to be produced in the interactions of
the UHE cosmic-rays (UHECR) with the CMB (i.e.  the GZK
process)~\citep{g,zk,kotera}.  The sources of the UHECR have not yet been identified, and it is
unknown if the sources are transient in nature or steady-state. Typical
GZK interaction lengths of a few hundred Mpc imply cosmogenic neutrinos will
retain the source direction over cosmological distances, but  any time
association with potential astrophysical transients is likely lost due to
deflections of UHECR by intergalactic magnetic fields. 

\emph{Astrophysical} neutrinos, believed to be produced directly in astrophysical
sources, have been detected at TeV-PeV energies by
IceCube~\citep{icecubeAstrophysical}.   IceCube has identified evidence for some
particular astrophysical neutrino sources, including \txs~\citep{txs_flare,txs_burst} and
NGC 1068~\citep{icecubeSource}.  Astrophysical neutrinos may also exist at UHE
energies, either as a continuation of the same flux that IceCube has detected,
or from other sources, such as flat-spectrum radio quasars (FSRQs)~\citep{fsrq1, fsrq2, Rodrigues_2018} or
GRBs~\citep{grb,grb2,grb3,grb4,grb5}. 

Compared to a diffuse UHE neutrino search, a search for UHE neutrinos
associated with particular sources can narrow the detection phase space in
direction and, for transient objects, in time. This in general allows a reduction
in backgrounds and/or an improvement in analysis efficiency, therefore
increasing the sensitivity compared to diffuse fluxes. 

In this paper,  we build on an ANITA-III diffuse search to develop a
methodology to search for UHE neutrinos in spatial and time coincidence with
astrophysical source classes.  We define a source class as a specification of
the time-dependent neutrino flux from one or more sources, $ \mathbf{F}(E,t)=
\sum_{\rm{sources}} F_i(E,t)$.  This methodology is applied to the ANITA-III
flight for five source classes: \txs, NGC 1068, blazars flaring in UHE
gamma-rays as identified by the Fermi All-sky Variability Analysis (FAVA~\citep{fava}), GRBs, and supernovae (SN).

\section{Simulation}

The standard ANITA simulation, \icemc~\citep{icemc}, is designed for efficient
simulation of a diffuse flux of neutrinos. The volumetric sampling
method used is efficient in sampling neutrinos likely to trigger ANITA, but is
not appropriate for modeling point sources, as it relies on the ``thin
target" approximation, converting effective volume to effective area by dividing
by the interaction length.  As such, a specialized sampling scheme was developed for
this search within \icemc. 

The first step is to choose a payload position/time and neutrino direction. A
random time is chosen within the ANITA-III flight which determines the
payload position and orientation.  The neutrino direction is chosen based on
the simulated parameters, for example, a single source, an isotropic flux, or from a
collection of time-varying sources. 

\begin{figure}[t]
  \centering
  \includegraphics[width=3.25in]{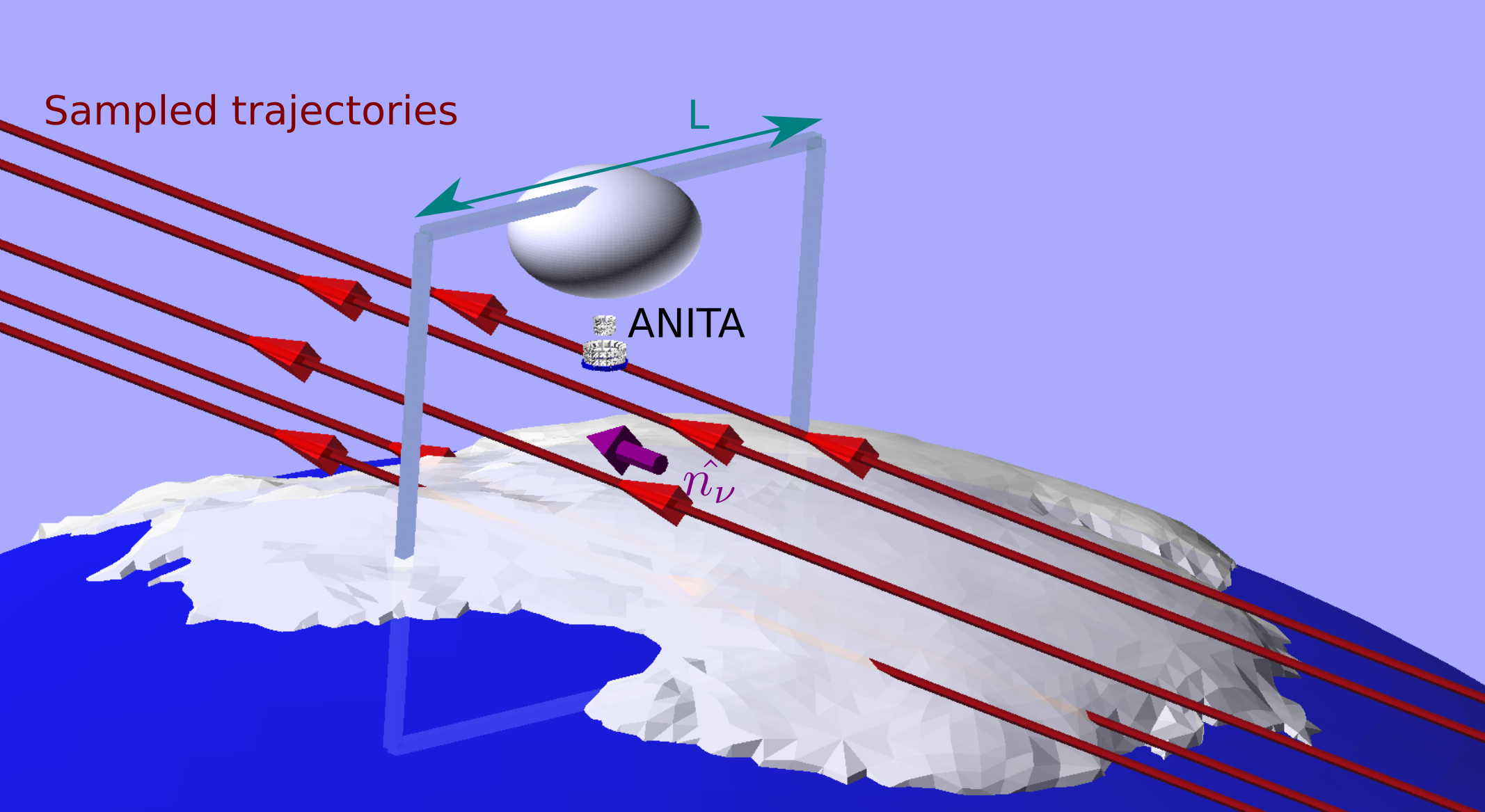} 
  \caption{ An illustration of the sampling method used for the point-source simulation. See text for details. } 
  \label{fig:sim}
\end{figure}

For a payload position, $\vec{x}_{\rm{ANITA}}$,  and neutrino direction, 
$\hat{p}_{\nu}$, one straightforward way to calculate the effective area for a
given neutrino energy, $E_{\nu}$, is
to consider a square with side $L$, normal to $\hat{p}_{\nu}$ and uniformly select a point on the square to define 
the neutrino trajectory (see Figure ~\ref{fig:sim}). Assuming that
the square is centered near $\vec{x}_{\rm{ANITA}}$ (we center it at the ice
surface below ANITA) and $L$ is made large enough so all neutrinos that could trigger ANITA 
intersect the sampling square, the trigger-level effective area for that configuration is then: 

\begin{equation}
  A_{\rm{eff}} (\hat{p}_{\nu}, \vec{x}_{\rm{ANITA}}, E_{\nu}) =  L^2 \frac{n_{\rm{trigger}}}{n_{\rm{thrown}}}.
\end{equation}
\noindent
A value of $L=1200$ km is a conservative choice that will miss no triggered neutrinos for ANITA-III. 

This scheme can be generalized to calculate ANITA's effective area to a point
source with sky position $\Theta$ by choosing the appropriate
$\hat{p}_{\nu}$ at each sampled $\vec{x}_{\rm{ANITA}}$, integrating over the flight trajectory by uniformly choosing random times within the flight: 

\begin{equation}
  A_{\rm{eff}} (\Theta,E_{\nu}) = L^2 \frac{n_{\rm{trigger}}}{n_{\rm{thrown}} }.
\end{equation} 

\begin{figure*}[t] 
  \centering 
  \textbf{Mean ANITA-III trigger-level point-source effective area}\\

  \begin{tabular}{cc} 
    Equatorial & Local \\ 
    \hspace{-0.5in}
  \includegraphics[width=0.47\textwidth]{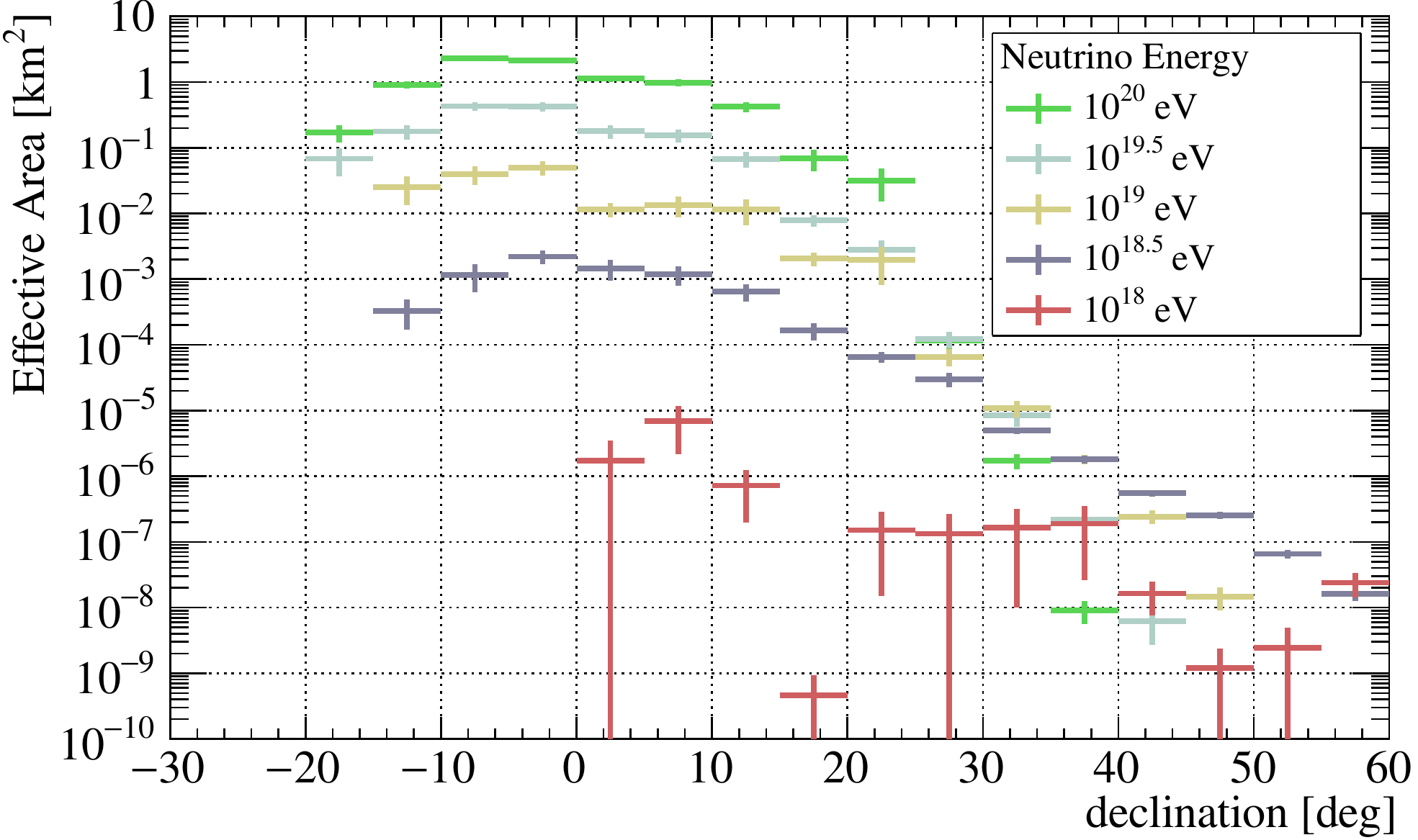} 
    & 
  \includegraphics[width=0.47\textwidth]{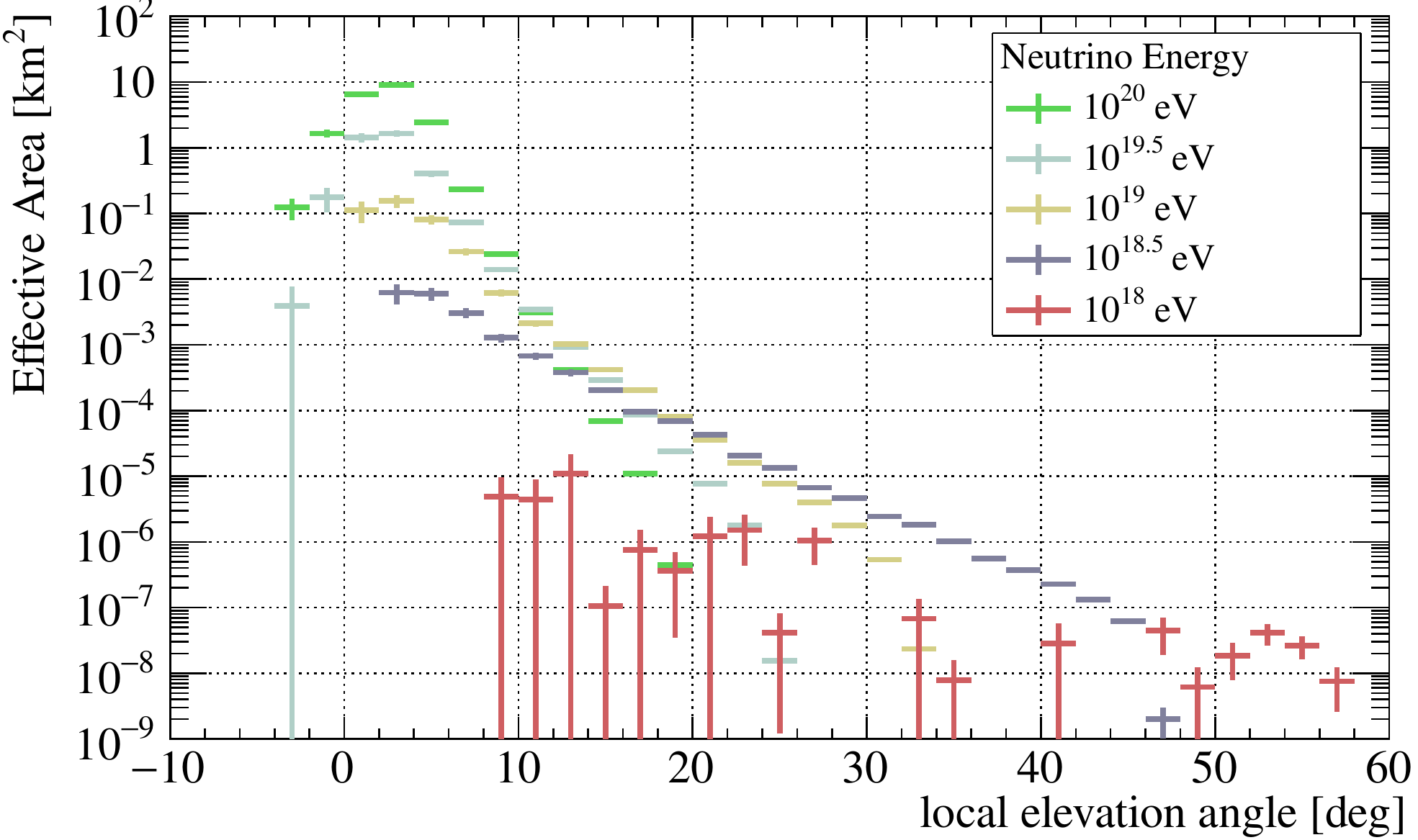} 
  \end{tabular}

  \caption{
    The point-source effective area as a function of declination (left) or
    local elevation angle (right) averaged over the ANITA-III flight at several
    energies.  Local elevation angle is
    measured downward from payload horizontal.  Because ANITA moves slowly
    compared to the diurnal cycle, the flight-averaged acceptance is nearly
    constant with right ascension. The total flight time is 22 days. 
    Errors shown are statistical only (systematic uncertainties are estimated to be of order 50\%).
    The
    point-source simulation results shown here are not directly comparable to the
    diffuse sensitivities quoted elsewhere by ANITA, as those diffuse sensitivities are
    geometric averages of ANITA's diffuse acceptance given by two independent simulation codes. 
  }
  \label{fig:dec}
\end{figure*} 

The remaining task is to calculate $n_{\rm{trigger}}$. The
obvious forward calculation---propagating a neutrino through the Earth
until it interacts or exits and then checking if ANITA would have triggered on interacting
neutrinos---requires a lot of computing time to acquire a sufficient sample of triggering neutrinos. To speed up calculation, we use the following scheme to calculate whether or not a given neutrino triggers ANITA: 

\begin{enumerate} 
  \item Check if the neutrino trajectory intersects an ellipsoid 5 km bigger than the geodesic reference ellipsoid (the highest altitude in Antarctica is 4,892 m).  If not, there is no chance of detection, so skip to the next event. 
  \item Find the intersection points of the neutrino with the Antarctic ice volume, which we take from BEDMAP2~\citep{BEDMAP2}). A step size of 50 m is used in order to detect intersections with ice at least that small. If no intersection with ice is detected, then there is no chance of detecting this neutrino, so skip to the next event. 
  \item A trajectory may have multiple intersections with the ice. For each intersecting segment, we calculate a weight, $w_s$, that is the product of the probability of the neutrino interacting within ice segment and the probability that the neutrino did not interact in the Earth prior to reaching the segment. 
  \item We choose one intersecting segment at random and pick an interaction point within that segment exponentially distributed from the entry point according to the cross-section. We correspondingly multiply $w_s$ by the number of intersecting segments to compensate for the selection. 
  \item We then proceed with the rest of the simulation as implemented in \icemc~(pick interaction type from differential cross-section, sample inelasticity, etc.) and evaluate if ANITA triggers on the radio emission. 
  \item $n_{\rm{trigger}}$ is the sum of $w_s$ for events that trigger. 
\end{enumerate}

This method can also be used to simulate a diffuse flux by choosing the source
direction at random in each trial, effectively integrating over $d\Omega$. By doing so, we can compare to the traditional
\texttt{icemc} sampling method and find that the diffuse effective areas agree
for ANITA-III at the 20\% level, which we consider an acceptable level of
agreement, subdominant to other systematic errors in the
simulation of order 50\%~\cite{icemc}. The time-averaged point-source UHE neutrino effective
areas as a function of declination and elevation angles over the ANITA-III
flight are shown in Figure~\ref{fig:dec}. 

Finally, this method may be adapted to simulate ANITA's
response to a source class.  We integrate over the flight trajectory by sampling times during the flight. At each time, we draw 
from the sources active at the time with probability proportional to its relative flux compared to
all other active sources and apply an additional time-dependent weight $w_t =   \mathbf{F}(E,t) /
\int_{t_0}^{t_1} \mathbf{F}(E,t) dt$.

\section{Source Search Methodology} 

We adapt Analysis B from the ANITA-III diffuse search, which is described
in detail in the appendices of~\cite{anita3}. A brief review is provided here. 

\subsection{Review of diffuse  ANITA analysis}

The ANITA payload consists of 48 dual-polarization horn antennas sensitive to a frequency range of approximately 200-1200 MHz. Whenever ANITA triggers, an
event is formed from 96 100-ns-long waveforms digitized from 48
dual-polarization antennas with known relative positions and time delays. These
waveforms are filtered to remove narrow-band contamination, then an
interferometric map is generated for each polarization, where the mean
cross-correlation between antennas is computed as a function of elevation and
azimuth in payload-centric coordinates. Directions corresponding to the peaks
of these maps are considered plane-wave source hypotheses and coherently-summed waveforms are
created in those directions, from which various observables are computed.
Analysis B applies three sets of cuts to select diffuse neutrino and air shower
candidates:

\begin{enumerate} 
  \item Quality Cuts $(\mathcal{Q})$, to remove digitizer glitches, radio interference from the payload itself, and other problematic pathologies. 
  \item A Fisher discriminant, $(\mathcal{F}$), selecting for neutrino-like events. This multivariate discriminant is constructed using a variety of observables from event waveforms (e.g. cross-correlation, coherence,  signal size, impulsivity, linear polarization) and is trained on a sideband of thermal events and simulation, to select impulsive broadband events. 
  \item A spatial isolation parameter, $\mathcal{O}$, based on projecting events to the continent, to remove likely anthropogenic events. This parameter is equal to the overlap integral of each event's pointing resolution projected onto the continent with that of other events. Only events with sufficiently high $\mathcal{F}$ are included in this integral, with a soft turn-on (so that events less likely to be neutrino-like are given a smaller weight). As $\mathcal{O}$ quickly becomes very small for events that are far apart, it is convenient to set cuts on $-\log_{10} \mathcal{O}$ instead of $\mathcal{O}$ directly. 
\end{enumerate}

Emission from neutrinos at the payload is typically vertically-polarized due to
the usual skimming geometry of UHE neutrinos and preferential transmission
at the ice-air interface. Conversely, radio emission from extensive air showers produce impulsive signals that are predominantly horizontally-polarized~\cite{anita_cr}. To avoid this background, only predominantly vertically-polarized events are
considered Askaryan neutrino candidates.  $\mathcal{Q}$ was optimized to reduce the
number of poor-quality events to a negligible level and then $\mathcal{F}$ and
$\mathcal{O}$ each had cut values optimized for sensitivity to a diffuse flux.
The primary contribution to the background estimate is anthropogenic, with
negligible background from thermal noise,  glitches, and payload-generated radio
interference.

\subsection{$\mathcal{D}$, the source class distance}

\begin{figure*}[t]
  \includegraphics[width=\textwidth]{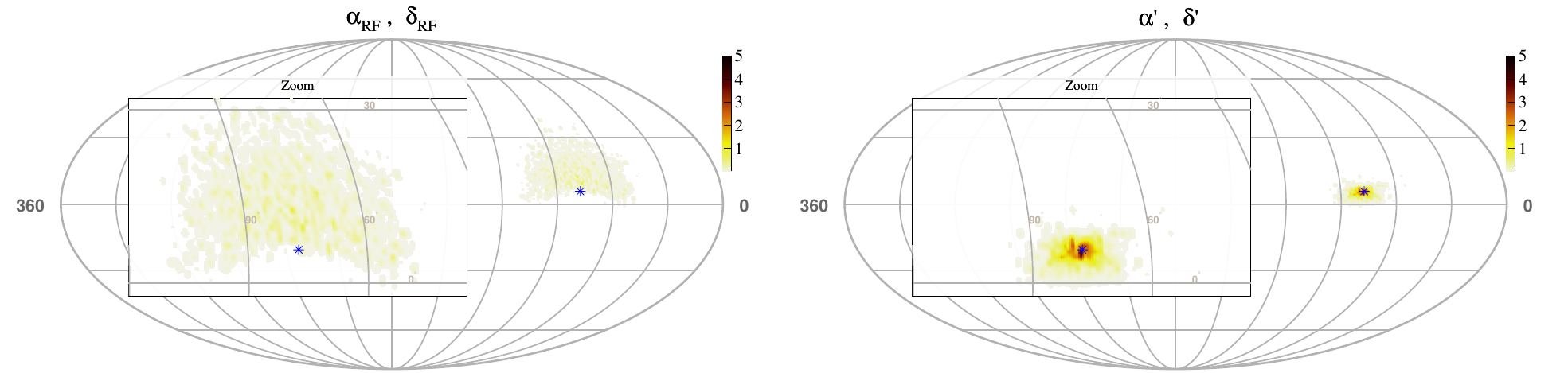} 

  \caption{
    The distribution in equatorial coordinates of the apparent RF direction
    (left) for simulated neutrinos from \txs, and the proxy sky position(right)
    computed using the approximate reconstruction method described in the text.
    The more compact distribution allows for improved background
    discrimination. The color axis depicts the percentage of spatial
    distribution within each angular bin ($1^{\circ} \times 1^{\circ}$); bins contributing less than 0.01
    percent are not shown for clarity. As one measure of the improvement in
    localization, the standard deviation in $\alpha$ improves from 13.3 degrees
    to 4.5 degrees and the standard deviation in $\delta$ improves from 6.4
    degrees to 1.7 degrees.
  } 

  \label{fig:dists} 
\end{figure*} 

We now develop a methodology to optimize the search for sensitivity to a given
source class rather than a diffuse flux.  The approach developed here strays
from the well-established likelihood-based methodology employed, for example, in
source searches by IceCube Observatory~\cite{icecubeSource}. That technique,
where a signal and background model are constructed and a likelihood analysis
is performed scanning over possible signal models, does not translate straightforwardly
to ANITA. The primary backgrounds to neutrino searches in  ANITA are not ``physical backgrounds" but
instead  anthropogenic radio emission. We don't know how to model this background and so cannot easily incorporate it into a likelihood analysis (we rely on
sidebands outside of the signal region to estimate it).  Moreover, we wish to
recycle most of the work from the diffuse search, which did not use a
likelihood-based methodology.

For expedience, we will keep the definitions of  $\mathcal{Q}$, $\mathcal{O}$, 
and $\mathcal{F}$ the same as in the diffuse search, but introduce a new
``source distance" parameter, $\mathcal{D}$  (to be defined shortly), that is a
measure of how closely-associated an event is to a source class.  We will then
optimize cut values for $\mathcal{F}$, $\mathcal{O}$ and $\mathcal{D}$ to
maximize search sensitivity for each source class. The $\mathcal{Q}$ cut values are kept the same as
in the diffuse analysis. 

The interferometric event reconstruction used by ANITA produces an estimated
direction of the radio emission ($\phi_{RF}$, $\theta_{RF}$) relative to the
payload. Since the orientation, time, and position of the payload at each event
are known, this direction may  be projected into equatorial coordinates, right
ascension ($\alpha$) and declination ($\delta$).  Using simulation, the
distributions of $(\alpha_{RF},\delta_{RF}$) may be estimated for a particular
neutrino flux. However, due to the opening angle of the in-ice emission cone,
these distributions are not compact for a single neutrino direction (see Figure
\ref{fig:dists}, left). In order to efficiently select neutrinos from a
direction, large swaths of the sky must be accepted, leading to a relatively
higher amount of background. As such, we desire an observable related to
neutrino direction that is more compact, which will tend to have higher
background rejection for a given signal efficiency. 

Reconstructing the neutrino direction for an event, rather than the direction
of radio emission, would meet this goal.  Because Askaryan emission is
radially-polarized, the observed polarization state of the radio emission from
neutrino is related to the neutrino direction.  Similarly, the power spectral
density of an event is related to how far away the emission angle is from the
Cherenkov angle~\cite{arz}. However, these observables are muddled by
instrumental and radio propagation effects, and moreover, the highly-varying
differential directional acceptance to neutrinos must be taken into account. 

For a putative ANITA neutrino candidate, a straightforward (but
dependent, as an energy spectrum must be assumed) way of estimating the
neutrino direction is to simulate neutrinos from different directions with the
payload fixed at the observation location in order to determine the neutrino
directions compatible with observables.  While this exercise would
be worthwhile for a sufficiently-interesting candidate and would result in a
compactly-defined unbiased neutrino direction distribution, it would be infeasible to perform this
procedure on the many events considered in a search.

Fortunately, for our purposes, we can live with an approximate reconstruction
of the neutrino direction, as any imperfections or biases would manifest themselves in both data and simulation.  To this end, we
derive a data-driven approximate reconstruction using machine learning
techniques. A large set of isotropically-distributed neutrinos were simulated
at various energies and run through the same reconstruction framework as data.

\begin{figure}[t]
  \centering
  \includegraphics[width=5in]{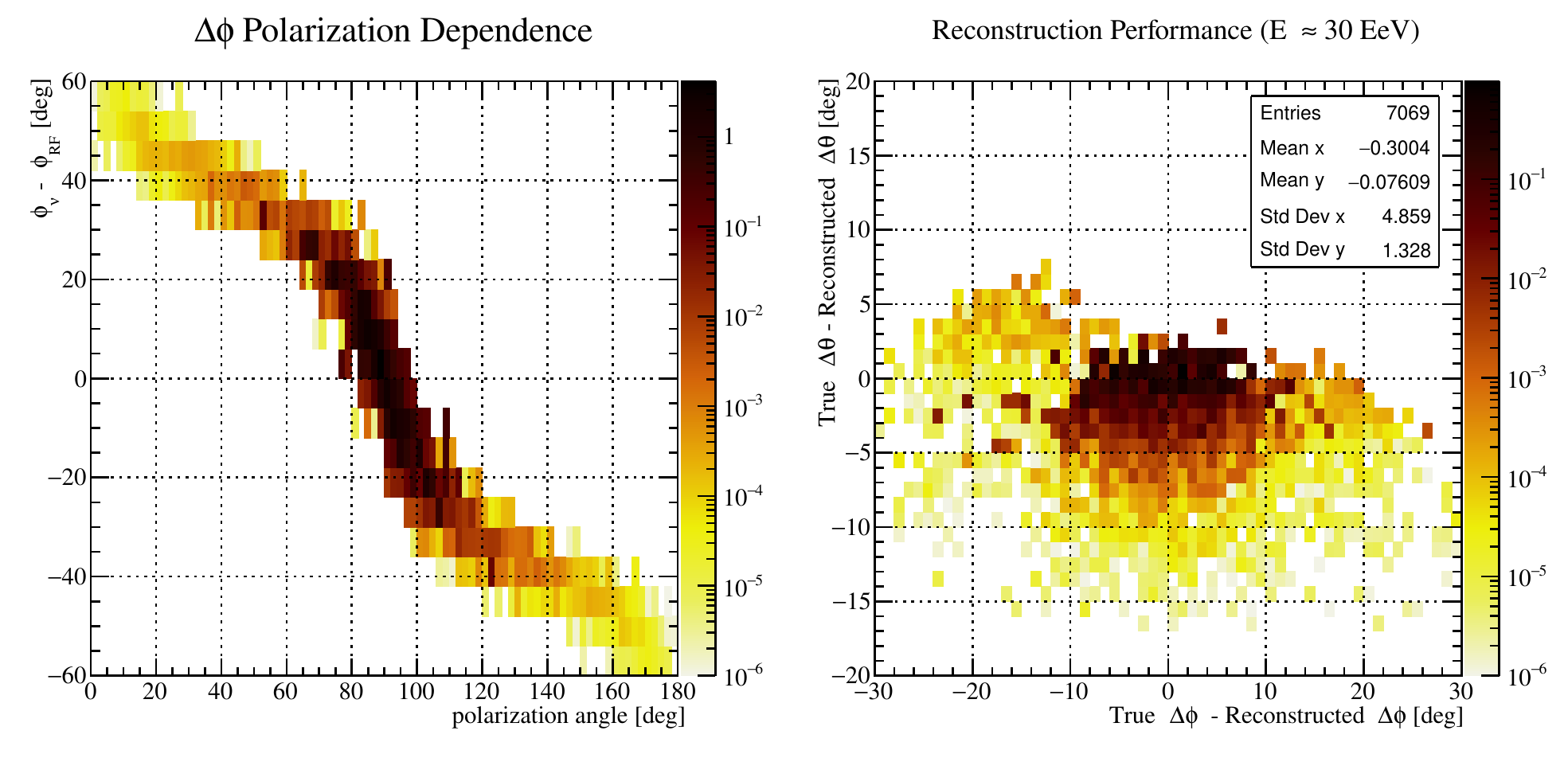} 

\caption{Left: The relationship between the reconstructed radio emission
  azimuth and the true neutrino direction as a function of reconstructed
  polarization angle for a diffuse neutrino flux. The reconstructed
  polarization angle is one component of the polynomial regression used for the fast approximate reconstruction. 
  Right: The accuracy of the reconstruction for $\Delta \phi$ and $\Delta \theta$ for neutrinos at 30 EeV. Note that the color axes, which represent the relative number of weighted neutrinos, are logarithmic. The majority of poorly-reconstructed events have very low weights. 
  }
  \label{fig:estimator} 

\end{figure}

TMVA~\citep{TMVA} was used to regress a polynomial for each of  $\Delta \phi =
\phi_\nu - \phi_{RF}$ and $\Delta \theta = \theta_\nu - \theta_{RF}$ as a
function of a number of observables derived from the waveforms that appeared to be related to the relative neutrino direction.  Fig.~\ref{fig:estimator} (left) shows, as an example,
the relationship between the reconstructed polarization angle and $\Delta \phi$ for the
training dataset. 

The
polynomials used were lightly hand-tuned until acceptable results were produced. Further tuning or more sophisticated
machine learning techniques could potentially yield even better results. 
The polynomials regressed were 
\begin{equation} 
  \Delta\phi = P_5(\theta_{pol})+ P_1(\theta_{RF})
\end{equation}
and
\begin{equation}
  \Delta\theta = P_5(\theta_{pol})+ P_1(\theta_{RF}) + P_3(m_{spec}) + P_3(BM) ,
\end{equation}
where $P_n$ denotes a polynomial of order $n$,  $\theta_{pol}$
is the reconstructed polarization angle (using the Stokes Parameters derived
from  dedispersed coherently-summed  waveforms of both polarizations),
$m_{spec}$ is an estimate of the spectral slope, and $BM$ is a measure of
occupied bandwidth of the coherently-summed waveform (similar to a Gini coefficient). 
For the final regression, we used a cosmogenic-like spectrum with additional events
added in the range 1-10 EeV, but the performance did not appear to depend vary much with the choice of spectrum. 

By applying this regression to each event, we can compute an estimator for
$\Delta\phi$ and $\Delta\theta$.  The performance of this estimator for
simulated diffuse events is shown in Fig.~\ref{fig:estimator} (right).  With
knowledge of the payload position and orientation, this can then be projected
to form our approximate equatorial coordinate estimator, or \emph{proxy sky position} ($\alpha',\delta'$).  We can test
the improved compactness of the proxy sky position for a source by applying the
reconstruction to Monte Carlo truth, as shown in Figure~\ref{fig:dists}, right.
We find an improvement in compactness for the proxy sky position of a factor of 3-4 in
each of $\alpha'$ and $\delta'$.  

\begin{figure}[b] 
  \centering
  \includegraphics[width=0.7\textwidth]{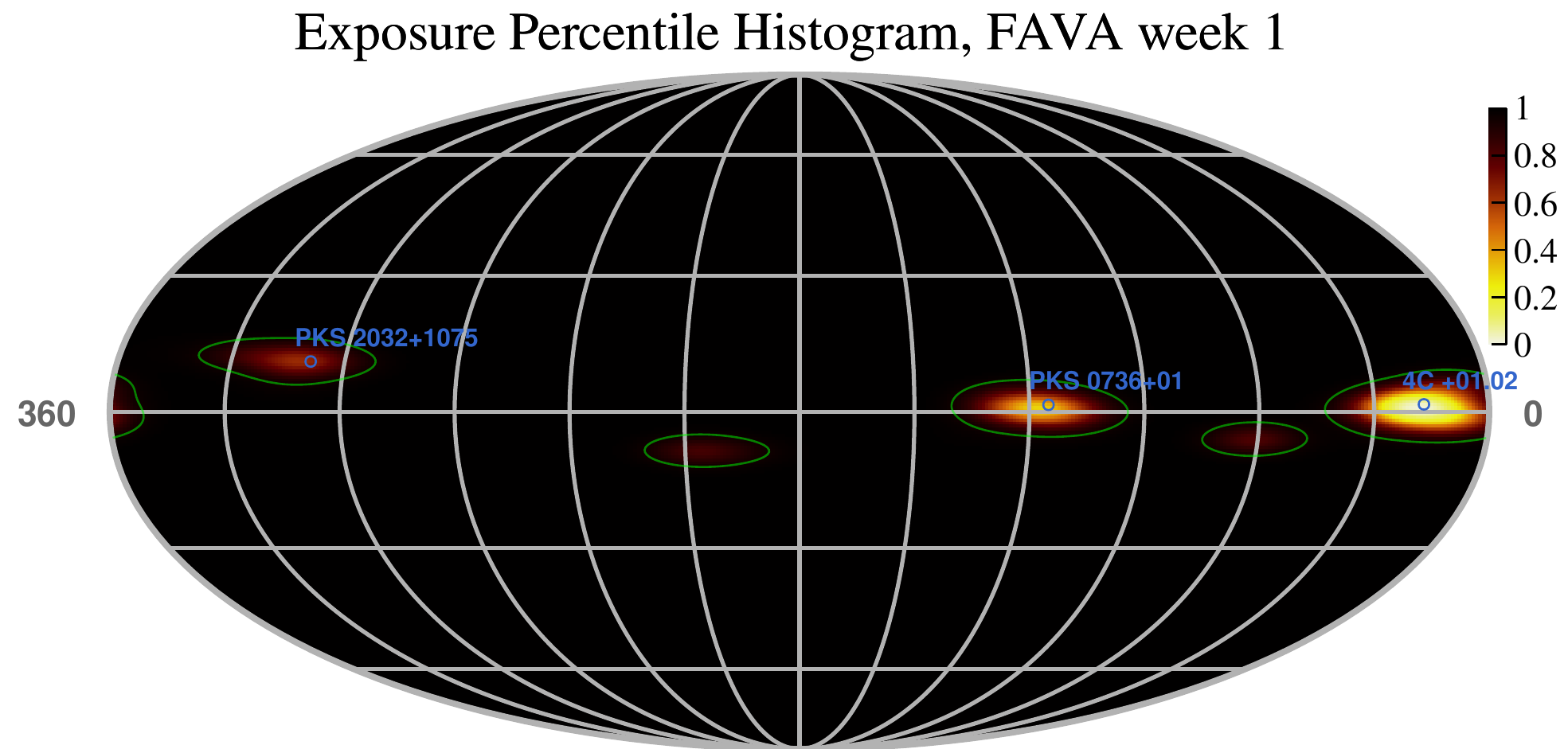} 

  \caption{An example time slice of an \emph{exposure percentile histogram}, in
  this case for the first week of FAVA-detected flaring blazars, which contains multiple sources of varying exposure. The value at each point $(\alpha',\delta',t)$ is set to the fraction of 
  the total exposure with exposure density higher than the exposure density of the bin, so that lower values represent regions of relatively greater exposure. A cut in
  $\mathcal{D}$ defines the region in $(\alpha',\delta',t)$  with value smaller
  than the cut value, preferentially selecting regions with the greatest
  exposure density. The sources corresponding to the three highest-exposure regions are labeled, although note that in general there may be a systematic offset between the proxy sky position and the actual source coordinates.
  The green contours correspond to $\mathcal{D}=0.96$.  
  }

  \label{fig:percentile} 
\end{figure}

It remains to convert our per-event proxy sky position,
$(\alpha',\delta')$, to a single source distance parameter to cut on.
To avoid any assumptions about the shape or modality of the distribution, we
use a simulation-driven approach to define $\mathcal{D}$.  For each
source class, we run a dedicated set of simulations with the appropriate
neutrino directions and spectra,  process the simulated events, and run the
estimator regression. We then produce a three-dimensional \emph{exposure
histogram} with axes of reconstructed $\alpha'$, $\delta'$, and event time.
For a steady source class, the time axis is extraneous and may be ignored. In
the case of discrete source turn-on and turn-off times, it is convenient to
choose the time bins to align with transitions.  

The value of each exposure histogram bin can be interpreted as the relative exposure (\emph{exposure density}) from this
source class for events at that reconstructed time and estimated direction. We seek to define a parameter that smoothly selects the highest exposure density regions. 
To do this, the exposure histogram is
converted to an \emph{exposure percentile histogram}, where each bin's
value is set to the fraction of the total neutrino exposure with exposure
density greater than the exposure of the bin (Figure
~\ref{fig:percentile}). This results in a histogram where the bins with the
highest exposure to the source class have values closer to 0 and bins with poor
or no exposure have a values close to 1.  We finally define $\mathcal{D}$ for
an event and source class  as the value of a source class's exposure percentile histogram for its $\alpha'$,
$\delta'$ and event time. This definition has the convenient property that
setting the cut on $\mathcal{D}$ to $d\in[0,1]$ selects the
highest exposure density region with an efficiency of approximately $d$. 

Because it is
time-consuming to generate enough simulated events to smoothly fill a
exposure histogram at fine resolution, we actually start with a relatively coarsely binned $\alpha'$
and $\delta'$ axes and use a Gaussian kernel density estimator in each time slice to
approximate a higher-resolution map. The kernel density estimator scale
parameters are tuned so that the distribution of $\mathcal{D}$ for members of
the source class is roughly uniform. Deviations from uniformity are not  
very concerning as long as efficiency varies smoothly with a cut on 
$\mathcal{D}$, so that the parameter can be effectively scanned. To reduce bias, disjoint sets
of simulated data are used to create the exposure percentile histogram and for cut optimization,
described next. 

\subsection{Cut Optimization} 

For each source class, we seek to set optimal cuts in a blind way on the $\mathcal{F}$
(the signal-likeness parameter), $\mathcal{O}$ (the spatial isolation
parameter), and $\mathcal{D}$ (the source class distance parameter). We scan in
these parameters to optimize expected sensitivity, using: 

\begin{equation} 
  \rm{sens.} \propto \left< \frac{FC_{90}\left(Pois\left(\mu_{BG}\right), \mu_{BG}\right) }{ \epsilon} \right>, 
\end{equation} 

\noindent
where $\mu_{BG}$ is the background estimate over the entire flight for a given set
of cuts, $\epsilon$ is the expected analysis efficiency for a given set of
cuts (the fraction of events that trigger ANITA that also pass analysis cuts), Pois is the Poisson distribution and FC$_{90}$(sig,bg) is the
90\% Feldman-Cousins upper limit factor~\citep{fc} for a given number of signal
events and background events. As both $\mu_{bg}$ and
$\epsilon$ have uncertainties, we calculate the expectation value in a semi-Bayesian way ~\citep{cousins-highland} by integrating over their posterior distribution.

The estimate for analysis efficiency for a given set of cuts can 
be estimated from simulation by applying the cuts to simulated data.
A Gaussian systematic error of 10\% on the analysis efficiency is assumed, as
derived in the diffuse analysis from calibration pulser data. 

To estimate the background as a function of cut values, we use a data-driven
on-off~\citep{LiMa} approach based on time shuffling.  For each event passing a trial
$\mathcal{F}$ and $\mathcal{O}$, we create 100 off-time pseudoevents by shifting the event time by a random
offset in either direction between 1.5 hours and 22.5 hours. These
bounds are chosen to guarantee  a significantly different sky position 
while preserving some time locality.  We consider the pseudo-events an off-time sideband approximately seven times larger in phase space than the ``on-time" signal region. 
We count the number of psuedoevents that pass the cut on $\mathcal{D}$ and the posterior on the background estimate is then conservatively taken to be the
Poisson posterior using a uniform prior for a sideband seven times larger: 

\begin{equation} 
  p(\mu_{BG}) = \Gamma( 7 N_{pass}/100 +1, 1/7). 
\end{equation} 

\noindent
A different choice of prior (e.g. Jeffreys) would reduce the background estimate at the cost of some non-conservative coverage. 
This method is limited by the statistics of the sideband; as soon as the cuts are made so stringent
that $N_{pass}$ is always zero, the background estimate will take a minimal value of ($0.10^{+0.16}_{-0.07}$), 
no matter how much more the cuts are tightened. Consideration of the efficiency
in the sensitivity will typically shift the optimization to the boundary,
somewhat alleviating this problem. An alternative would require a different strategy such as
imposing a background model, which is difficult for the time-varying
anthropogenic backgrounds faced by ANITA. 

Using this methodology, we can perform a three-dimensional scan over reasonable
parameters of $\mathcal{F}$,  $-\log_{10} \mathcal{O}$, and $\mathcal{D}$ to
choose cuts that optimize our sensitivity metric for any particular source
class.  Finally, as in the diffuse search,  we only select events that are more
impulsive in vertical polarization (VPol) than horizontal polarization (HPol)
to avoid selecting air shower events. 

\section{Sources considered }

Having developed a methodology to optimize cuts for a source class, we now
turn our attention to potentially interesting sources for ANITA-III. Sources of
the same type are pooled together into a single source class in order to reduce
the global trials factor. This requires some model dependence in choosing the
analysis cuts, but once cuts are chosen, model-independent limits may be placed
on each source within the class. These limits may not be optimal for any given
model other than the one used to set cuts, but we make an attempt to adopt a
``least-common denominator" set of cuts to minimize the model-dependence. 

Each search is optimized and performed independently, but the significance of
any particular search's result must be interpreted in the context of the number
of searches performed.  We considered optimizing for a global 90\% significance
(which is roughly equivalent to setting a higher optimized significance in each
search, if they are weighted equally), but found that this did not strongly
affect where the optimized cuts are placed, and we did not want to restrict the
possibility of any additional future searches. It is also possible for a given
event to be considered a candidate by multiple searches. The considered objects
are tabulated in Table~\ref{tbl:sources} and the result of each 
optimization is shown in Table~\ref{tbl:efficiency}.

\begin{table*}[p] 
  \begin{center}

    \small

    \textbf{Objects Considered}
  \begin{tabular}{|c|c|c|c|} 
	\hline 
    Object & Search & Coordinates & Times Considered (UTC) \\
    \hline
		\hline
    \txs & \txs & $\alpha=77.4^{\circ}$, $\delta=5.7^\circ$ & Full Flight\\
    \hline
    NGC 1068 & NGC 1068 & $\alpha=40.7^{\circ}$, $\delta=-0.0^\circ$ & Full Flight\\
    \hline
     3C 454.3 & Flaring Blazar & $\alpha=344^\circ,\delta=16.1^\circ$ & 2014-12-15-15:43:38Z + 1 week \\
     4C +01.02 & Flaring Blazar & $\alpha=17^\circ,\delta=1.6^\circ$ & 2014-12-15-15:43:38Z + 4 weeks \\
     \rm{*} B3 1343+451 & Flaring Blazar & $\alpha=206^\circ,\delta=44.8^\circ$ & 2015-01-05-15:43:38Z + 1 weeks \\
     CTA 102 & Flaring Blazar & $\alpha=331^\circ,\delta=11.7^\circ$ & 2014-12-22-15:43:38Z + 3 weeks \\
     MG1 J221916+1806 & Flaring Blazar & $\alpha=335^\circ,\delta=18.0^\circ$ & 2014-12-15-15:43:38Z + 2 weeks \\
     \rm{*} PKS 0402-362 & Flaring Blazar & $\alpha=61^\circ,\delta=-36.0^\circ$ & 2014-12-15-15:43:38Z + 4 weeks \\
     PKS 0502+049 & Flaring Blazar & $\alpha=76^\circ,\delta=5.0^\circ$ & 2014-12-22-15:43:38Z + 3 weeks \\
     PKS 0736+01& Flaring Blazar & $\alpha=115^\circ,\delta=1.5^\circ$ & 2014-12-15-15:43:38Z + 2 weeks \\
     PKS 1441+25& Flaring Blazar & $\alpha=221^\circ,\delta=25.0^\circ$ & 2014-12-15-15:43:38Z + 4 weeks \\
     PKS 1717+177& Flaring Blazar & $\alpha=260^\circ,\delta=17.7^\circ$ & 2014-12-22-15:43:38Z + 2 weeks \\
     \rm{*} PKS 1830-211& Flaring Blazar & $\alpha=278^\circ,\delta=-21.1^\circ$ & 2015-01-05-15:43:38Z + 1 weeks \\
     PKS 2032+1075& Flaring Blazar & $\alpha=309^\circ,\delta=10.9^\circ$ & 2014-12-15-15:43:38Z + 1 weeks \\
     \rm{*} PKS 2052-47& Flaring Blazar & $\alpha=314^\circ,\delta=-47.3^\circ$ & 2014-12-22-15:43:38Z + 2 weeks \\
     \rm{*} PKS 2142-75& Flaring Blazar & $\alpha=327^\circ,\delta=-75.7^\circ$ & 2014-12-15-15:43:38Z + 1 weeks \\
     PKS B1319-093& Flaring Blazar & $\alpha=200^\circ,\delta=-9.3^\circ$ & 2014-12-15-15:43:38Z + 1 weeks \\
     \rm{*} PMN J2141-6411& Flaring Blazar & $\alpha=325^\circ,\delta=-64.2^\circ$ & 2014-12-29-15:43:38Z + 1 weeks \\
     RGB J2243+203& Flaring Blazar & $\alpha=341^\circ,\delta=20.3^\circ$ & 2014-12-15-15:43:38Z + 2 weeks \\
     \rm{*} S4 +1144+40& Flaring Blazar & $\alpha=177^\circ,\delta=40^\circ$ & 2014-12-22-15:43:38Z + 1 weeks \\
     \rm{*} S5 +1217+71& Flaring Blazar & $\alpha=185^\circ,\delta=71.1^\circ$ & 2014-12-15-15:43:38Z + 2 weeks \\
     TXS +1100+122& Flaring Blazar & $\alpha=166^\circ,\delta=12.0^\circ$ & 2014-12-29-15:43:38Z + 2 weeks \\
    \hline
     GRB 141221A & GRB &$\alpha=198.3^\circ,\delta=8.2^\circ$& 2014-12-21-08:07:02Z $_{\mathrm{- 5 min}}^{\mathrm{+ 1 day}}$ \\
     \rm{*} GRB 141223240 & GRB &$\alpha=147.4^\circ,\delta=-20.7^\circ$& 2014-12-23-05:45:34Z $_{\mathrm{- 5 min}}^{\mathrm{+ 1 day}}$ \\
     GRB 141226880 & GRB &$\alpha=163.9^\circ,\delta=28.4^\circ$& 2014-12-26-21:07:24Z $_{\mathrm{- 5 min}}^{\mathrm{+ 1 day}}$ \\
     GRB 141229911 & GRB &$\alpha=170.1^\circ,\delta=23.1^\circ$& 2014-12-29-21:51:39Z $_{\mathrm{- 5 min}}^{\mathrm{+ 1 day}}$ \\
     \rm{*} GRB 141229A & GRB &$\alpha=72.4^\circ,\delta=-19.2^\circ$& 2014-12-29-11:48:59 $_{\mathrm{- 5 min}}^{\mathrm{+ 1 day}}$ \\
     GRB 141230834 & GRB &$\alpha=181.5^\circ,\delta=11.6^\circ$& 2014-12-30-20:00:25Z $_{\mathrm{- 5 min}}^{\mathrm{+ 1 day}}$ \\
     GRB 141230A & GRB &$\alpha=57.0^\circ,\delta=1.6^\circ$& 2014-12-30-03:24:22Z $_{\mathrm{- 5 min}}^{\mathrm{+ 1 day}}$ \\
     GRB 150101B & GRB &$\alpha=188.0^\circ,\delta=-10.9^\circ$& 2015-01-01-15:23:00Z $_{\mathrm{- 5 min}}^{\mathrm{+ 1 day}}$ \\
     GRB 150105A & GRB &$\alpha=124.3^\circ,\delta=-14.8^\circ$& 2015-01-05-06:10:00Z $_{\mathrm{- 5 min}}^{\mathrm{+ 1 day}}$ \\
     GRB 150106921 & GRB &$\alpha=40.8^\circ,\delta=0.3^\circ$& 2015-01-06-22:05:53 $_{\mathrm{- 5 min}}^{\mathrm{+ 1 day}}$ \\
    \hline

    \rm{*} SN 2014dz & SN & $\alpha=52.1^\circ,\delta=38.0^\circ$ & 2014-12-10-00:00:00Z  + 2 weeks\\
    SN 2014dy & SN & $\alpha=42.2^\circ,\delta=-0.8^\circ$ & 2014-12-10-00:00:00Z  + 2 weeks\\
    \rm{*} SN 2015A & SN & $\alpha=145.3^\circ,\delta=35.9^\circ$ & 2015-01-02-00:00:00Z  + 2 weeks\\
    SN 2015B & SN & $\alpha=193.6^\circ,\delta=-12.6^\circ$ & 2014-12-21-00:00:00Z  + 2 weeks\\
    SN 2015D & SN & $\alpha=198.2^\circ,\delta=12.6^\circ$ & 2015-01-06-00:00:00Z + 2 weeks\\
    SN 2015E & SN & $\alpha=48.4^\circ,\delta=0.3^\circ$ & 2014-12-31-00:00:00Z + 2 weeks\\
    SN 2015W & SN & $\alpha=104.4^\circ,\delta=13.6^\circ$ & 2015-01-02-00:00:00Z + 2 weeks\\
    \hline
  \end{tabular}
  \end{center} 
  \caption{All objects considered in this search, along with coordinates and times the source is assumed to be turned on. ANITA-III launched Dec 18, 2014 and was terminated Jan 9, 2015.  Objects with asterisks were included in the search but resulted in no passing simulated events. The times for the flaring blazar search are one-week time scales based on the Fermi telescope's elapsed mission time, hence the offset of 15:43:38 for each start time.}
  \label{tbl:sources}
  \end{table*}

\begin{table}[tp]
  \begin{center}
    \textbf{Optimized cut values}\\
  \footnotesize
\begin{tabular}{|l|c|c|c|c|c|}
  \hline
  Search & $\mathcal{F}  $ & $-\log_{10} \mathcal{O}  $ & $\mathcal{D}  $ &  exp. $\epsilon$ & exp. backg.\\
\hline
  & & & & & \\[-8pt]
  \txs & $\geq$ 1.7 & $\geq$ 0.9 &$\leq$ 0.99 & 91.7\%&  $0.10^{+0.16}_{-0.07}$\\ [3pt]
  NGC 1068& $\geq$ 2.8 & $\geq$ 0.5 &$\leq$ 0.99 & 93.9\% & $0.10^{+0.16}_{-0.07}$\\ [3pt]
  FAVA blazars& $\geq$ 2.3 &$\geq$  0.9 &$\leq$ 0.96 & 82.2\%  & $0.30^{+0.25}_{-0.16}$ \\ [3pt]
  GRBs& $\geq$ 1.7 & $\geq$ 0.1 & $\leq$ 0.96 & 92.7\%& $0.10^{+0.16}_{-0.07}$ \\ [3pt]
  SN& $\geq$ 2.3 &$\geq$ 0.9 & $\leq$ 0.99 &  93.0\%& 0$.23^{+0.22}_{-0.13}$ \\ [3pt]
\hline
\end{tabular}
  \end{center}
  \normalsize
\caption{The optimized efficiency and background for each search performed for ANITA-III. The quoted efficiency is calculated for the model used to optimize the cuts and has an estimated uncertainty of order 10\%. The background is model-independent. The 16\%, 50\% and 84\% quantiles of the posterior distribution are used to determine the central value and errors quoted. }
\label{tbl:efficiency} 

\end{table}

\subsection{\txs} 

The \txs~ blazar has been identified by IceCube as a potential source of
astrophysical neutrinos. This association is due to a gamma-ray flare in spatial and temporal
coincidence with a September 22, 2017 likely astrophysical neutrino candidate
that triggered a multi-messenger alert~\citep{txs_flare}.  Afterwards, archival
data suggested an excess of neutrinos from the direction of \txs~in a
several-month window around December 2014~\citep{txs_burst}, albeit without any
gamma-ray activity in the blazar~\citep{txs_no_gamma}. The ANITA-III flight coincided temporally
with this earlier  neutrino ``burst" and \txs, at a declination of 5.6
degrees, is within ANITA-III's sensitive field-of-view, motivating a dedicated
search. 

IceCube has measured a spectral index for the neutrino burst of
$\gamma=2.1\pm{0.2}$. To optimize cuts for the ANITA-III search, we simulate
neutrinos from the direction of \txs~with $\gamma=2$, which is compatible with
the IceCube measurement and somewhat preferred theoretically.  The optimization
results in an estimated analysis efficiency to an $E^{-2}$ flux of 91.7\% with
a background estimate of $0.10^{+0.16}_{-0.07}$, which is the minimal
background estimate calculable with the method employed (i.e. zero passing
sideband events).

\subsection{NGC 1068} 

IceCube has identified NGC 1068 as a potential neutrino point
source~\citep{icecubeSource} at the 2.9$\sigma$ level. IceCube does not report
any temporal information for NGC 1068, and the best-fit spectral index (3.2) from
IceCube would make detection by ANITA-III unlikely. However, as it is one of
just two objects within ANITA's field of view that has been identified as a
potential high-energy neutrino source at this significance level, a search is
performed.  

As the best-fit spectral index of $\gamma=3.2$ would make detection by ANITA virtually impossible, and moreover, there is no
guarantee that a source must have consistent spectral index over
many orders of magnitude of energy, we set cuts by simulating neutrinos from NGC 1068
with $\gamma=2$, resulting in an estimated analysis efficiency of 93.9\% and a background of $0.10^{+0.16}_{-0.07}$ (the minimal allowed).

\subsection{Flaring Blazars} 

Motivated in part by the apparent \txs~flare coincidence, we consider
blazars within the field of view of ANITA-III that are flaring in GeV
gamma-rays as potential UHE neutrino sources. FSRQs in particular have been suggested as particularly-efficient neutrino sources above 1 EeV~\citep{fsrq1,fsrq2, Rodrigues_2018}, although we consider all classes of blazars in this search. We use the Fermi Large Aperture
Telescope's All-sky Variability Analysis (FAVA)~\citep{fava} to select flaring
objects labeled as blazars by 3FGL~\citep{3fgl} during the ANITA-III flight.  FAVA identifies flaring candidates
on a one-week cadence, during which we assume the neutrino flux is constant. 

To set cuts, we weight neutrino flux from each blazar equally. While weighting
by gamma-ray flux or luminosity distance are also reasonable, equal weighting
is the least model-dependent, making it less likely to miss any interesting
source. As before, we assume $\gamma=2$. The result of the optimization is an analysis efficiency estimate of 82.8\% and an estimated background of $0.30^{+0.25}_{-0.16}$.

\subsection{GRBs} 

GRBs, the brightest known transient events in the universe,  have long been considered a potential source of UHE neutrinos~\citep{grb,grb2,grb3,grb4,grb5}.
ANITA is most likely most sensitive to the GRB afterglow neutrino flux, which
is expected to have $\gamma=3/2$ up to some maximum energy (model-dependent,
but typically order EeV). %

We select GRBs from the IceCube GRBWeb catalog~\citep{icecubeGRB}, which itself
combines data from several sources~\citep{swiftGRB,fermiGRB}.  To
optimize cuts, we chose to adopt a $\gamma=3/2$ spectrum, typical of afterglow models, up to 10 EeV for each
GRB, starting five minutes before the GRB and extending 24 hours after. While
most models would not predict neutrinos up to 10 EeV for most GRBs, this choice is
inclusive and will avoid missing any potential signals.  This time window would
also tend to accept prompt and precursor neutrinos. The period of 24 hours is
chosen as a compromise between too short a window, which might reject some of the most
energetic afterglow neutrinos, and too long a window, which increases the
possibility of chance coincidence. For the purpose of cut optimization, the relative normalization of each source's flux was
assumed to be proportional to the GRB fluence as measured by the Fermi Gamma-ray Burst Monitor, using a reasonable typical value if it was not available. Only GRBs with a declination
within 30 degrees of the equator are considered. The cut optimization indicates
an estimated analysis efficiency of 92.7\% and the minimal possible background estimate of
$0.10^{+0.16}_{-0.07}$.

\subsection{Supernovae} 

Phenomena related to supernovae (SN) have been predicted to produce UHE neutrinos, especially in cases such as Hypernovae, magnetar-driven SN, transrelativistic SN, or tidal ignition of white dwarfs~\citep{SN_HN, SN_TR, SN_MG, SN_TD1,SN_TD2, SN_nearby}. Despite the lack of a clear model that might produce an observable signal in ANITA, their transient nature makes them amenable to a
source search. Furthermore, the upward-air shower candidate in ANITA-III was
spatially coincident with Supernova 2014dz \citep{a3upward}, which likely
occurred only several days before, further motivating a search in the Askaryan
channel. 

Due to a lack of clear model guidance for setting cuts, we select $\gamma=2$ and
a two-week period after the estimated explosion date of the supernova, which is generally computed using spectral properties of the light curve at discovery. We select SN
from the CBAT catalog~\citep{cbat} and do not distinguish between supernova
types. Tidal disruption events (TDEs) were also considered, but none were catalogued near the time of the ANITA-III flight~\citep{tde_catalog}. Optimization of cuts results in an estimated analysis efficiency of 93.0\% on a background of $0.22^{+0.22}_{-0.13}$.

\section{Results and Discussion} 

\begin{figure*}[tp]
\includegraphics[width=\textwidth]{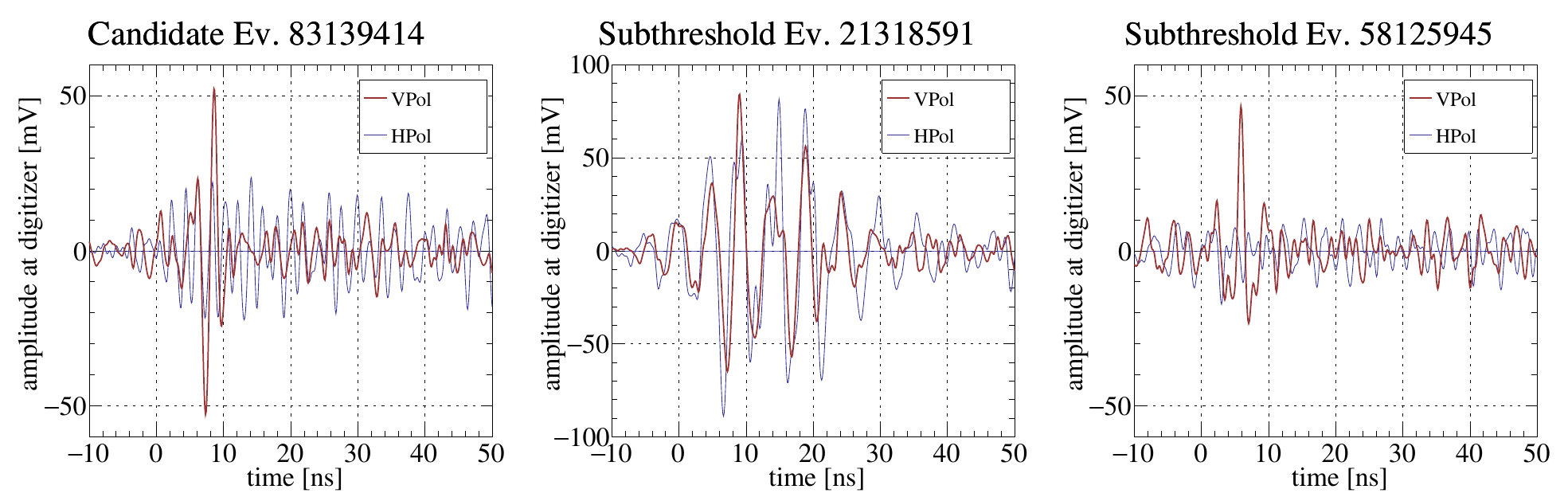} 
  \caption{The candidate event (left) and two subthreshold events identified in this analysis. These waveforms represent a coherent sum based on the estimated radio direction and with the group delay from filters and antennas removed.} 
  \label{fig:events} 
\end{figure*}

After applying the optimized cuts to each search using the procedure described above, all searches were null except
for the SN search, which identified event 83134914 (Figure~\ref{fig:events}, left) as potentially associated
with SN 2015D ($\mathcal{D} =0.67$). This is consistent with the background
estimate for the search ($p=0.21$), even before accounting for the number of
searches performed. 

Due to a bookkeeping error in background estimation, initially a
looser set of optimized cuts were computed and erroneously applied, which resulted in two additional
passing events, 21318591 and 58125945 (Figure~\ref{fig:events} center and right). 
As this initial unblinding did not follow the procedure prescribed ahead of time and these subthreshold events do not pass the final, corrected cuts, we do not consider them part of the result, though we briefly comment on them. 
All three events are summarized in Table~{\ref{tbl:events}}. 

\begin{table*} 

  \centering
    \textbf{Details about identified events}\\[1em]

    \small
    \begin{tabular}{|c|c||c|c|} 
    \hline
            & \textbf{Candidate}    &    Subthreshold & Subthreshold \\
         & \textbf{Ev. 83134914}     &   Ev. 21318591 & Ev. 58125945  \\ 
    \hline
    \hline
    Time & 2015-01-08-19:04:24.237&   2014-12-22-04:30:24Z & 2015-01-01-08:17:14.615Z \\ 
    \hline
    Payload Pos. & 70.3S, 90.1E, 33.6 km & 80.2S, 82.0E, 36.6 km & 76.2S, 108.6W, 34.6 km \\ 
    \hline
    Est. Ice Pos.. & 68.6S, 98.2ES  & 81.8S, 94.7E & 74.4S, 100.3W  \\ 
    \hline
    Proxy Sky Pos.  &$\alpha'=206^\circ$,$\delta'=13.6^\circ$& $\alpha'=38^\circ$, $\delta'=-3.1^\circ$&  $\alpha' = 164^\circ, \delta'=11.7^\circ$ \\ 
    \hline
    Potential & SN 2014 D ($\mathcal{D}=0.67$)& 4C +01.02  ($\mathcal{D}=0.955)$ &  TXS 1100+122 ($\mathcal{D}=0.003)$  \\
    Associations && NGC 1068 ($\mathcal{D}$ =0.64)  & \\ 
       && SN 2014dy ($\mathcal{D}$ =0.78) &   \\ 
    \hline
    $\mathcal{F} $ & 3.03 & 2.28 & 3.06  \\ 
    \hline
    $-\log_{10} \mathcal{O}$ & $\sim\infty$ & 0.41 & 0.68 \\ 
    \hline
  \end{tabular} 

  \caption{Details about the candidate event and subthreshold events that were associated with searches. Event 83134914 was identified by the SN search as potentially associated with SN 2015D. This event was previously identified in the diffuse ANITA-III analysis. The other two events were subthreshold and revealed due to an error in the initial unblinding, but are included here for completeness.}

  \label{tbl:events} 
\end{table*}

\subsection{Candidate Event 83139414} 

The sole candidate event this search, event 83134914, was previously identified in the ANITA-III diffuse
search~\citep{anita3} as being neutrino-like and extremely isolated. In this
search, it was found to be potentially associated with SN 2015D~\citep{cbet4051}, which was
discovered approximately 10 days after this event
and believed to be around two weeks old at the time of discovery.  

We note that that the proxy position for this event in this search
($\alpha'=206$~,~$ \delta'=13.9$) differs substantially from what was reported
in the diffuse search ($\alpha=171\pm5$~,~$\delta=16.3\pm1$), a position which would have
failed the association cut here. The previous estimate was performed by
dedicated simulations with the payload at that position, rather than the
approximate method used in this paper, which is designed to reduce dispersion rather than produce a bias-free position estimator. Moreover, the handling of polarization in
the simulation software has also been improved since the previous estimate was
made, adding another potential source of discrepancy. For the purpose of this search, the
event is associated even if the localization proxy differs from the previous
estimate. 

This event points to isolated deep ice, very far from any known anthropogenic
activity and would have passed a much more stringent isolation cut. If, hypothetically, the selection cuts
were set to barely accommodate this event, the background estimate for the
supernova search would be the minimal possible with the method used
($0.10^{+0.16}_{-0.07}$), which would modify the pre-trial significance of this
event to $p=0.13$. This is not enough to be interesting by itself, but is perhaps more interesting in combination with 
the potential association of the apparently upward air-shower in
ANITA-III with an SN, which had $p=0.0017-0.023$, depending on the prior used for the time dependence~\citep[Supplemental Material]{a3upward}.

To get a feeling for the false association rate, should 83134914 be an
anthropogenic event, we consider the sideband of VPol-identified events with
$\mathcal{F} \in [1,2]$, $-\log_{10} \mathcal{O} \in [-1,1]$. Of these 173
sideband events, 4 (2.3\%) would have passed the SN association cut
value and just one event has an association as good as the candidate. 22 events
(12.7\%) in this sideband pass the association cut for any of the
five source classes.

It is possible that 83134914 is a UHE neutrino event and not some
anthropogenic background, but is associated with a SN direction by
chance. By simulating a diffuse cosmogenic neutrino flux we find 10\% of simulated diffuse
neutrinos would be considered SN-associated in this search and 24\% would be
considered associated with any of the source classes at each search's cut level.

A subthreshold search in the direction of SN 2015D down to $\mathcal{F}=0$ and
$-\log_{10} \mathcal{O}=0$ was performed, yielding one additional event. This event was
isolated but was only marginally-associated with SN2014D and upon further
inspection, is potentially residual radio interference from satellites. 

\subsection{Subthreshold Events 21318591 and 58125945}

The cuts associated with the initial unblinding of the searches
erroneously underestimated the background estimate due to a software bug involving an inconsistency in histogram binning.  This resulted in less stringent optimized cuts that selected  events 21318591 and 58125945 as candidates. 
We stress that these events do not pass the final cuts after correctly applying the optimization procedure and are not considered a part of the results. However, these events would
represent the result of a valid---albeit less sensitive---search with slightly
higher efficiency and a significantly higher (by a factor of 2-3) background
estimate. We briefly discuss these accidentally unveiled events
here for completeness and transparency. 

Event 21318591 was initially considered a candidate in multiple searches (NGC
1068, SN, Flaring Blazars), although, with $\mathcal{F}=2.3$ and $-\log_{10}
\mathcal{O} = 0.4$,  it fails the corrected cuts for all searches, is less
impulsive than a typical neutrino event, has an unlikely polarization for a
neutrino, and the nearby events have similar shapes. 

Event 58125945 was initially identified as associated with flaring blazar TXS 1100+122.  With
$-\log_{10} \mathcal{O} = 0.7$, it fails the corrected isolation
cut of 0.9 for the FAVA search, although it is the closest event to passing. This event is impulsive and virtually purely VPol, but points to
the Pine Island Glacier, a part of the continent with relatively
low likelihood to produce detectable Askaryan neutrino events due to high
temperature and relatively low ice thickness. While fairly isolated, the
nearest neighbors, which form a cluster with each other, look broadly similar
to this event.
Moreover, the British Antarctic Survey was conducting radar and drilling studies in Pine
Island Glacier during the 2014-2015 season~\citep{TraverseInformation}. 
Due to a storm on January 1, there was no significant activity at the time of
the event~\citep{mulvaney_personal_communication}, but a storm also 
admits the possibility of
triboelectric emission~\citep{experimentalSubsurface}. 
While this event is clearly consistent with background and did not pass the
corrected cut, we note that TXS 1100+122 has been suggested as a potentially
interesting neutrino source~\citep{txs1100_2020}, on the basis of TXS 1100+122
being compatible in position with IceCube alert event
(IC-200109A~\citep{IC200109A}) and possessing a compact radio emission core, a
feature suggested as possibly being associated with neutrino
emission~\citep{vlbi_nu}.

\subsection{Limits} 

As we do not find any significant associations, we proceed to set upper
limits on fluence during the ANITA-III flight from each source. Model-independent limits on fluence at a given energy $\Phi(E)$ are given by 
\begin{equation}
u.l. (E^2 \Phi(E)) = \left < \frac{FC_{90} ( n_{obs},  \mu_{bg}) \cdot E}{\epsilon A_{eff}(E) \cdot \Delta} \right> , 
\end{equation}

\noindent where $\Delta$ is a factor compensating for the use of discrete energies, while the flux and acceptance both evolve continuously.
ANITA cosmogenic searches
have typically used $\Delta=4$ \citep{icemc},  based on studies showing that this
is a reasonable choice for a variety of models~\citep{rice}. We maintain this convention here for consistency, but note that other experiments use different conventions. 

\begin{figure*}[p]
\includegraphics[width=\textwidth]{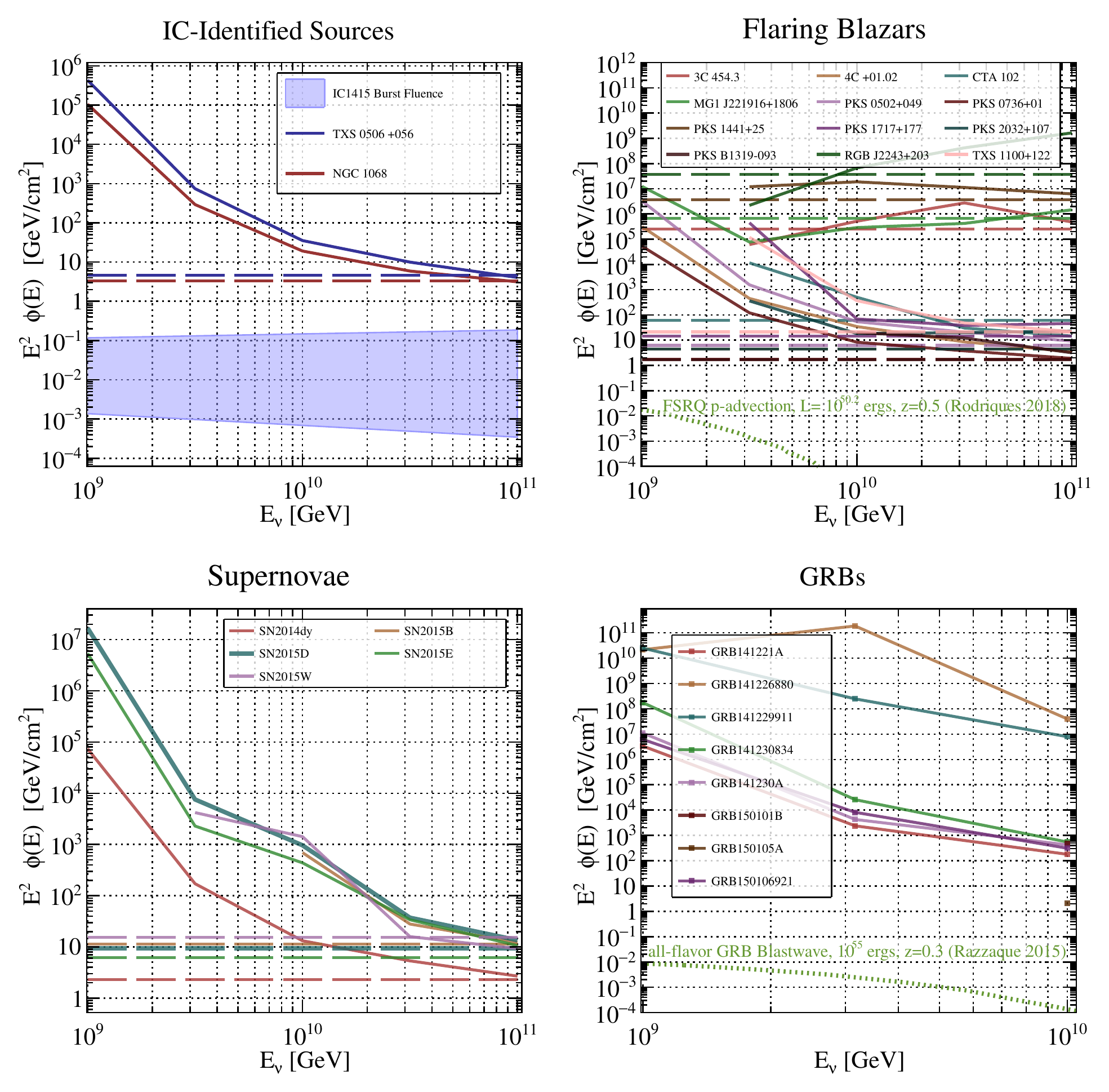} 
\centering
  \caption{ANITA-III limits on neutrino fluence for all objects considered in these searches. 
  The top-left panel shows fluence limits on \txs~and NGC 1068, along with the extrapolated time-scaled fluence from the 2014-2015 apparent IceCube neutrino burst in the direction of \txs. 
  The top right shows fluence limits flaring blazars that result in triggered simulated events (points are not shown at energies with no simulated passing events). Some blazars (e.g RGB J2243+203) suffer from poor geometry. We include an optimistic model for neutrinos from an FSRQ in a high-luminosity state, based on ~\citep{Rodrigues_2018}. The bottom left shows fluence limits on SN and the bottom right shows fluence limits on GRBs. For GRBs, we compare the limits to a model for neutrinos produced in the afterglows from a nearby, luminous GRB ~\citep{grb4}. 
  Model-independent limits are shown with solid lines and integrated $E^{-2}$ limits are shown with dashed lines, where appropriate.  }  
  \label{fig:limits}

\end{figure*}

Setting limits on objects within multi-member source classes requires additional
consideration in the handling of the background estimate,  as this estimate applies to the entire class, rather than
individual objects.  However, the true limit is be bounded by ascribing zero background to a source and
ascribing the total background of the source class to a source, and this
difference is small ($\mathcal{O}(10\%)$) in all cases here. We choose to ascribe zero background as that results in more conservative (higher) limits.  Upper limits for each
source are shown in Figure~\ref{fig:limits}. Some sources are omitted due to a
paucity of simulated triggered events or low analysis efficiency.

In addition to model-independent energy-dependent limits, we show integrated
$\gamma=2$ limits for AGN-like and SN searches, where the upper limit  on
normalization for a $\Phi(E) = \Phi{_0} (E^{-2})$ is calculated using: 

\begin{equation} 
  u.l. (\Phi_{0}) = \left <  \frac{FC_{90} ( n_{obs},  \mu_{bg}) \cdot \phi_0 }{ \left<\epsilon \cdot A_{eff}(E) \right>|_{\phi(E)}} \right>  ,
\end{equation} 

\noindent
where $A_{eff}$ and $\epsilon$ are averages over the unity-normalized flux $\phi(E) = \phi_0 E^{-2}$, with $\phi_0 = 1.00$ EeV for the energy range considered.

The measured fluence for the apparent 2014-2015 \txs~neutrino flare from
IceCube was $E^2 \Phi(E) = (0.21^{+0.09}_{-0.07})~\mathrm{GeV cm}^{-2}
(E/10^5 \mathrm{GeV})^{2-2.1\pm0.2}$ over a Gaussian window centered on 2014-12-14$\pm 14$
days with a width of $110^{+34}_{-24}$ days~\citep{txs_burst}. The ANITA-III flight time represents a fraction of 0.16 of this time window. This fluence
band,  scaled to the ANITA-III flight time, is projected onto the top-left panel of
Figure~\ref{fig:limits}.  We also superimpose some relevant models for GRBs and flaring blazars.

\section{Conclusion and Outlook} 

We find that there is no significant evidence for any source-associated
neutrinos with ANITA-III, although the potential SN association of the
ANITA-III diffuse analysis event is somewhat intriguing, especially in
combination with previous results. 

The methodology developed is shown to be capable of achieving higher efficiencies at lower backgrounds compared to a diffuse search. However, because the analysis efficiency of the
ANITA-III diffuse search was already high ($>$80\%), there is relatively little room for neutrinos that could have triggered
ANITA-III but not have passed analysis cuts employed in the diffuse search. As such, a null result is not surprising. Still, using this
methodology we are able to set limits on individual sources with ANITA, which can not be
done coherently in the diffuse search. A similar search is in progress for the
more recent and more sensitive ANITA-IV payload, although that also had a high diffuse analysis efficiency.

For similar experiments where the achieved diffuse analysis efficiency at an
acceptable background level is not as high, the methodology described here can
be more impactful, as there is additional phase space for discovery currently not
accessible to diffuse searches.  For example, 
a significant gap currently exists between diffuse analysis and trigger efficiency in
the Askaryan Radio Array (ARA) experiment~\citep{araPaper}. Employing  
an adaptation of the method described here therefore has the potential to discover new candidates within the ARA dataset 

Similarly, it follows that it is advantageous for future UHE radio neutrino
experiments to reduce trigger thresholds below expected achievable analysis
thresholds for diffuse searches. A reduced trigger threshold could not be
achieved in ANITA-III or ANITA-IV as the acquisition system could not handle
the corresponding increased data rate.  Future UHE neutrino detectors using the
Askaryan method such as PUEO~\citep{pueo}, RNO-G~\citep{rno-g}, or the radio
extension of IceCube Gen2~\citep{gen2radio} will use improved trigger
techniques capable of substantially reducing the achievable trigger thresholds
while maintaining lower rates. If the threshold is chosen to be low enough,
then the techniques developed here can potentially unveil candidate events not
discoverable in a diffuse search. 

The method outlined here suffers from an inability to reduce background below
some level, due to the use of a sideband for background estimation.  Finding
additional handles on background could potentially further reduce background and help improve sensitivity.
This may be accomplished through improved modeling of anthropogenic backgrounds
(easier in the case of fixed-position detectors than balloon payloads) or through the use of
sidebands with relatively more phase space (for example, by improving 
angular reconstruction or introducing additional variables). With the minimal
background estimate achievable in this analysis, 
four events are necessary for $>3\sigma$ evidence and eight events would be
required for $>5\sigma$ discovery, before accounting for trials factors. Doubling the relative size of the sideband,
for example, would reduce the number of events required to exceed each significance
threshold by one.  We note that the planned PUEO mission is projected to be more than an order of magnitude more sensitive than
ANITA-III, implying that a transient fluence producing one event in ANITA-III would likely yield a statistically-significant excess in PUEO. 

\acknowledgments

This work was supported by NASA grants NNX11AC44G, NNX15AC24G, and 80NSSC20K077.  We thank the staff of the Columbia
Scientific Balloon Facility for their generous support as well as logistical
support provided by the National Science Foundation and United States Antarctic
Program.  Additional support was provided by the Kavli Institute for Cosmological
Physics at the University of Chicago. Computing resources were provided by the
Research Computing Center at the University of Chicago and the Ohio
Supercomputing Center at The Ohio State University. A.  Connolly would like to
thank the National Science Foundation for their support through CAREER award
1255557 as well as grant 1806923. O. Banerjee and L. Cremonesi’s work was supported by collaborative
visits funded by the Cosmology and Astroparticle Student and Postdoc Exchange
Network (CASPEN). The University College London group was also supported by the
Leverhulme Trust. The National Taiwan University group is supported by Taiwan’s
Ministry of Science and Technology (MOST) under its Vanguard Program
106-2119-M-002-011. We thank Robert Mulvaney of the British Antarctic Survey for helpful information about BAS activities around the Pine Island Glacier. We additionally thank Kohta Murase and the anonymous referee for helpful suggestions.

We acknowledge use of the following software packages in this work: 
\texttt{ROOT}~\citep{ROOTsoft,ROOTpaper},
\texttt{TMVA}~\citep{TMVA}, 
\texttt{icemc}~\citep{icemc},
ANITA Collaboration Software~(\href{https://github.com/anitaNeutrino}{https://github.com/anitaNeutrino},~\cite{anita3}).

\bibliography{main}{}
\bibliographystyle{JHEP}

\end{document}